\documentclass[a4paper]{spie}  %>>> use this instead for A4 paper
\usepackage[]{graphicx}
\usepackage{amsmath}
\usepackage{amsfonts}
\usepackage{amssymb}
\usepackage{color}
\usepackage[pdftex]{hyperref}
\usepackage{subfigure}
\usepackage{geometry}
\title{Flaglets for studying the large-scale structure of the Universe}

\author{Boris Leistedt\supit{a}, Hiranya V. Peiris\supit{a}, and Jason D. McEwen\supit{a,b}\skiplinehalf
\normalsize{\supit{a}Department of Physics and Astronomy, University College London, London WC1E 6BT, U.K.\\
\hspace*{-2.5mm} \supit{b}Mullard Space Science Laboratory (MSSL), University College London, Surrey RH5 6NT, U.K.}}

\authorinfo{\{boris.leistedt.11, h.peiris, jason.mcewen\}@ucl.ac.uk}
\renewcommand{\exp}[1]{\ensuremath{{{\rm exp}({#1})}}}

\newcommand{\vect}[1]{\ensuremath{\mbox{\boldmath ${#1}$}}}

\newcommand{\eqn}[1]{Eqn.~{#1}}
\newcommand{\fig}[1]{Figure~{#1}}
\newcommand{\ball}{\mathbb{B}^3}
\newcommand{\sphere}{\mathbb{S}^2}
\newcommand{\sas}{\theta, \phi}
\newcommand{\sasprime}{\theta^\prime, \phi^\prime}
\newcommand{\ie}{i.e.}
\newcommand{\eg}{e.g.}
 
\begin{document} 
  \maketitle 

%%%%%%%%%%%%%%%%%%%%%%%%%%%%%%%%%%%%%%%%%%%%%%%%%%%%%%%%%%%%% 
\begin{abstract}
Pressing questions in cosmology such as the nature of dark matter and dark energy can be addressed using large galaxy  surveys, which measure the positions, properties and redshifts of galaxies in order to map the large-scale structure of the Universe. We review the Fourier-Laguerre transform, a novel transform in 3D spherical coordinates which is based on spherical harmonics combined with damped Laguerre polynomials and appropriate for analysing galaxy surveys. We also recall the construction of flaglets, 3D wavelets obtained through a tiling of the Fourier-Laguerre  space, which can be used to extract scale-dependent, spatially localised features on the ball. We exploit a sampling theorem to obtain exact Fourier-Laguerre and flaglet transforms, such that band-limited signals can analysed and reconstructed at floating point accuracy on a finite number of voxels on the ball. We present a potential application of the flaglet transform for finding voids in galaxy surveys and studying the large-scale structure of the Universe. 
\end{abstract}

%>>>> Include a list of keywords after the abstract 

%%%%%%%%%%%%%%%%%%%%%%%%%%%%%%%%%%%%%%%%%%%%%%%%%%%%%%%%%%%%%
\section{Introduction}\label{sec:intro} 

The large-scale structure of the Universe refers to the distribution of matter on cosmological scales, which contains a wealth of  information that can be used to test models of the origin, the content and the evolution of the Universe. In particular, galaxies and clusters of galaxies organise themselves in filaments and sheets which form a ``cosmic web"  carrying important signatures of the underlying physics. This cosmic web can be probed by analysing the distribution and properties of galaxies measured in large surveys of the sky. Modern surveys cover significant portions of the sky and reach faint luminosities, enabling us to map millions of galaxies, which trace the large-scale structure in three-dimensions over large cosmological volumes. However, this mapping is non-trivial due to selection effects and systematic uncertainties present in the data, for example resulting from spatial variations in the magnitude limits, the calibration of the instruments and the observing conditions over time. In addition, galaxies are not directly observed in three dimensions: their positions are complemented with distance information through their redshift, which is estimated from the spectra of their light. This yields data with spherical geometry and complex selection effects which complicate subsequent cosmological analyses. This setting requires the use of appropriate methods that can deal with the complexity of the data while probing the underlying three-dimensional large-scale structure. 

Currently, 3D analyses of galaxy surveys are performed using methods where angular and radial components are not separable, such as the Fourier-Bessel transform. Hence the descriptions of the selection effects and uncertainties in the data are non-trivial in these bases. In this work, we recall the definition of the Fourier-Laguerre transform, a 3D spherical transform constructed by Leistedt and McEwen\cite{leistedt:flaglets} which is fully separable and theoretically exact in both the continuous and discrete settings thanks to a sampling theorem on the ball. This property can be exploited to decompose and reconstruct band-limited signals at floating point accuracy in a finite number of voxels. Furthermore, the Fourier-Bessel transform of a signal which is band-limited in Fourier-Laguerre space can be calculated exactly, which allows one to use the convenient properties of the Fourier-Bessel basis, where the reconstruction of density and velocity fields is straightforward\cite{erdogdu20046df, erdogdu20062mass}.

The main features of the large-scale structure, such as filaments, sheets and voids, are spatially localised and occur at specific scales, characteristic of the underlying physics. Wavelets naturally extract such scale-dependent features and may thus provide a powerful tool to analyse large-scale structure data.  The flaglet transform introduced by Leistedt and McEwen\cite{leistedt:flaglets} is a novel wavelet transform built on the Fourier-Laguerre basis which allows one to probe scale-dependent features on the sky and along the line of sight separately. Therefore this transform can accommodate the geometry of galaxy survey data while being sensitive to localised structures in the cosmic web. In this paper, we highlight the potential of flaglets for finding cosmic voids in galaxy surveys. Voids are the large, underdense regions that occupy a large fraction of the volume of the Universe and are a natural consequence of the hierarchical growth of structure\cite{Weygaert2011voiddynamics, Aragon2013hierach,Sheth2004hierach,Colberg2005lcdmvoids}. They have a high potential for discriminating among models of dark energy and modified gravity\cite{BLi2013modgravityvoids,Sutter2012alcockvoids,Biswas2010voidsdarkenergy}, thanks to their statistical power and their quasi-linear gravitational regime (whereas galaxies are locally subject to strong non-linear gravitational effects). However, it is currently difficult to identify voids in a robust manner in galaxy surveys, and it is unclear how the complex selection effects and uncertainties present in the data may affect the cosmological analyses. Thanks to the properties of flaglets, a flaglet-based void finder has the potential to overcome these limitations and to be used to produce large catalogues of cosmic voids to study the large-scale structure of the Universe. 

The remainder of this article is organised as follows.  In Section \ref{sec:flaglets} we define the Fourier-Laguerre and the flaglet transforms on the ball, as well as the relation to the Fourier-Bessel transform. In Section \ref{sec:voids} we investigate the use of flaglets for finding cosmic voids in galaxy survey data. Concluding remarks are made in Section \ref{sec:summary}.

%%%%%%%%%%%%%%%%%%%%%%%%%%%%%%%%%%%%%%%%%%%%%%%%%%%%%%%%%%%%%
\section{Flaglets: Fourier-Laguerre wavelets on the ball}\label{sec:flaglets}

In this section we recall the definition of the Fourier-Laguerre transform\cite{leistedt:flaglets}, a separable harmonic transform in 3D spherical coordinates which combines the spherical harmonics on the sphere with damped Laguerre polynomials on the radial half-line. The Fourier-Laguerre transform is furthermore theoretically exact thanks to a sampling theorem on the ball, guaranteeing that band-limited signals can be decomposed and reconstructed exactly on a finite number of voxels. We then detail a tiling of the Fourier-Laguerre harmonic space with scale-discretised generating functions which yield wavelets with good localisation properties in real and harmonic space. The resulting flaglet transform\cite{leistedt:flaglets} can be used to analyse signals defined on the ball to extract scale-dependent features on the ball. We finally relate the Fourier-Laguerre approach to the canonical Fourier-Bessel transform and provide conversion formulae. 

%%%%%%%%%%%%%%%%%%%%%%%%%%%%%%%%%%%%%%%%%%%%%%%%%%%%%%%%%%%%%
\subsection{The Fourier-Laguerre transform}\label{sec:flag}

\begin{figure}
\centering
\subfigure[Basis functions {$K_p(r)$}]{\includegraphics[trim=7.5cm 23.2cm 1.5cm 0cm, clip, width=8cm]{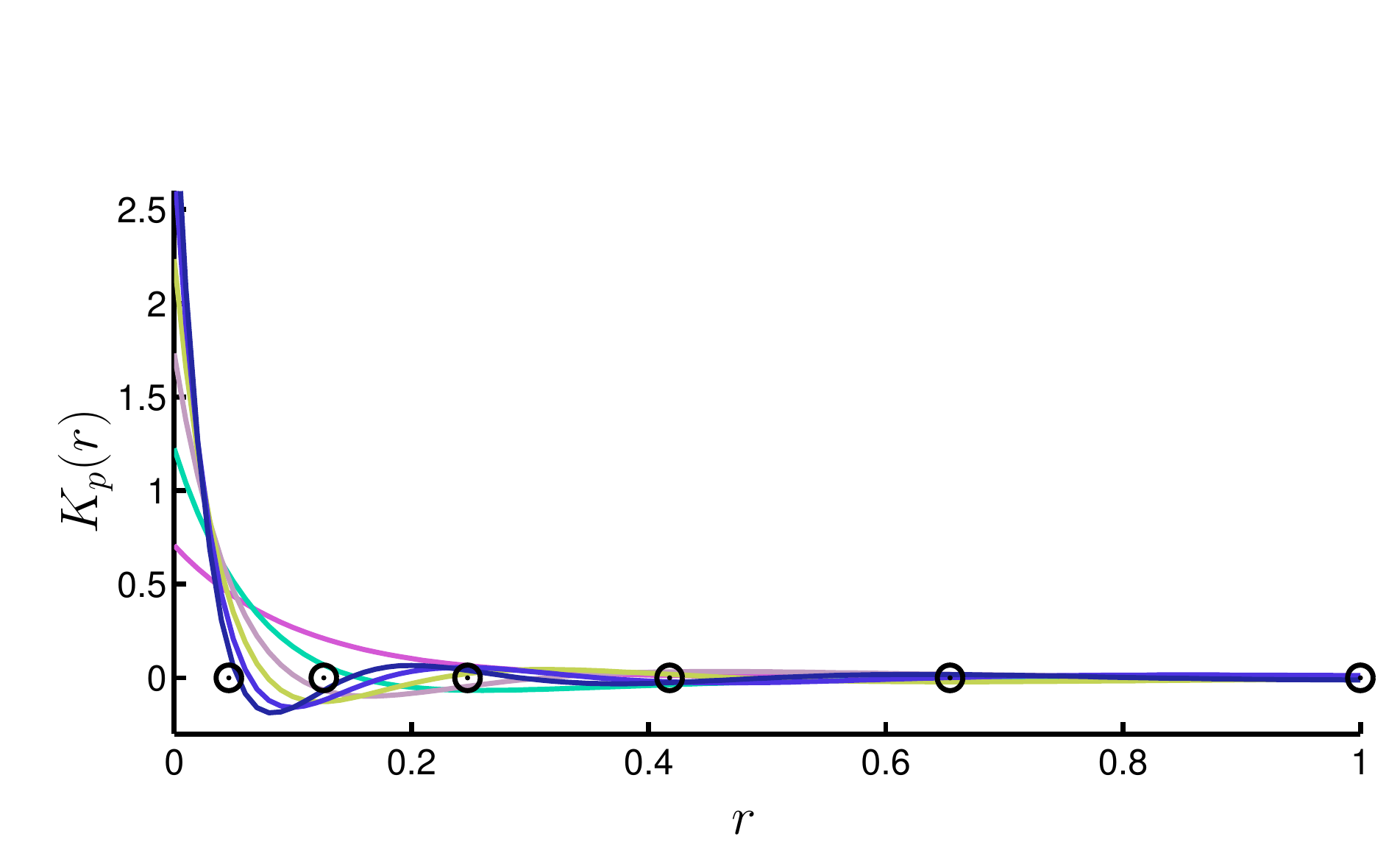}}
\subfigure[Zoom on the oscillatory features of {$K_p(r)$}]{\includegraphics[trim=7.5cm 12.7cm 1.5cm 12cm, clip, width=8cm]{pics/sphelaguerrebasis.pdf}}\\
\caption{First six spherical Laguerre basis functions $K_p(r)$ constructed on $r \in [0,1]$ and the associated sample positions (circles) obtained with the sampling theorem on the radial half-line (derived from Gaussian quadrature; see Ref.~\citenum{leistedt:flaglets}). In the context of the spherical Laguerre transform on the radial half-line\cite{leistedt:flaglets}, a function $f(r)$ with $r \in \mathbb{R}^+$ with band-limit $P=6$ can be decomposed and reconstructed exactly using these six basis functions only. In that case, $f$ and the basis functions are evaluated solely at the sampling points. }
\label{fig:fig1}
\end{figure}

To obtain a separable harmonic transform in 3D spherical coordinates, a straightforward approach is to combine transforms on the sphere and on the radial half-line. For the Fourier-Laguerre transform\cite{leistedt:flaglets}, we define basis functions on the ball {$\ball= {\mathbb{R}^+} \times \sphere$} such that 
\begin{equation}
  Z_{\ell m p}(\vect{r}) = K_{p}(r) Y_{\ell m}(\sas) ,
\end{equation}
with spherical coordinates \mbox{$\vect{r} = (r, \sas) \in \ball$}, where $r \in \mathbb{R}^{+}$ denotes radius, $\theta \in[0,\pi]$ colatitude and $\phi \in[0,2\pi)$ longitude, and where $\ell,p \in\mathbb{N}_0$ and $m\in\mathbb{Z}$ such that $|m|\leq\ell$. The standard spherical harmonics are denoted by $Y_{\ell m}$ and the normalised spherical Laguerre basis functions are defined on the radial half-line as
 \begin{equation}
	K_p(r) \equiv \sqrt{ \frac{p!}{(p+2)!} }  \frac{ e^{-{r}/{2\tau}} }{ \sqrt{\tau^3}} L^{(2)}_p\left(\frac{r}{\tau}\right) , % \quad {\rm with} \quad L_p^{(2)}(r) \equiv \sum_{j=0}^{p}  {p+2 \choose p - j} \frac{(-r)^j}{j!} 
\end{equation}
where $L^{(2)}_p$ is the $p$-th generalised Laguerre polynomial of order two and $\tau \in \mathbb{R}^+$ is a radial scale factor. These basis functions are orthonormal with respect to the measure $r^2 {\rm d} r $ and probe oscillatory features on the radial half-line. They form a complete basis, yielding the spherical Laguerre transform and a sampling theorem on the radial half-line derived from Gaussian quadrature, as detailed in Ref.~\citenum{leistedt:flaglets} and shown in \fig{\ref{fig:fig1}}.

The orthogonality and completeness of the 3D Fourier-Laguerre basis functions follow from the corresponding properties of the individual basis functions, where the orthogonality relation is given explicitly by the following inner product on $\ball$:
\begin{eqnarray}
	 \langle Z_{\ell m p} | Z_{{\ell^\prime} m^\prime p^\prime} \rangle_{\ball}
         &=& \int_{\ball} {\rm d}^3\vect{r} Z_{\ell m p} Z^*_{\ell^\prime m^\prime p^\prime} (\vect{r})  \ = \ \delta_{\ell \ell^\prime} \delta_{mm^\prime} \delta_{pp^\prime},
\end{eqnarray}
where ${\rm d}^3\vect{r} = r^2 \sin\theta {\rm d}r {\rm d}\theta {\rm d}\phi$ is the volume element in spherical coordinates. Any three-dimensional signal \mbox{$f \in L^2(\ball)$} can be decomposed as
\begin{equation}
	f(\vect{r}) = \sum_{p = 0}^{\infty}\sum_{\ell = 0}^{\infty}\sum_{m = -\ell}^{\ell} f_{\ell m p} Z_{\ell m p} (\vect{r}), \label{flaginverse}
\end{equation}
where the harmonic coefficients are given by the usual projection
\begin{equation}
	{f}_{\ell m p} = \langle f | Z_{\ell m p} \rangle_{\ball} =   \int_{\ball} {\rm d}^3\vect{r}  f(\vect{r}) Z^*_{\ell m p}(\vect{r}).  \label{flagforward}
\end{equation}
In what follows, we consider band-limited signals with angular and radial band-limits $L$ and $P$ respectively, \ie\ signals $f$ such that $f_{\ell m p} = 0$, $\forall \ell \geq L$, $\forall p \geq P$. In this case the summations in \eqn{\ref{flaginverse}} over $\ell$ and $p$ may be truncated to $L-1$ and $P-1$ respectively.

In practice, computing the Fourier-Laguerre transform involves the evaluation of the integral of \eqn{\ref{flagforward}}. An exact quadrature rule for the evaluation of this integral for a band-limited function $f$ naturally gives rise to a sampling theorem.  Since the Fourier-Laguerre transform is separable in angular and radial coordinates, we may appeal to separate sampling theorems on the sphere and radial half-line.  For the angular part, we adopt the equiangular sampling theorem on the sphere developed recently by McEwen and Wiaux\cite{mcewen:fssht} (MW).  Other sampling theorems on the sphere could alternatively be adopted (\eg\  the sampling theorem of Discoll and Healy\cite{driscollhealy1994}), however we select the MW sampling theorem since it leads to the most efficient equiangular sampling of the sphere (\ie\ the fewest number of samples to represent a band-limited signal exactly).  For the radial part, we used the sampling theorem developed for the spherical Laguerre transform, detailed in Ref.~\citenum{leistedt:flaglets}.  Combining these results we recover a sampling theorem and, equivalently, an exact Fourier-Laguerre transform on $\ball$.  For a band-limited signal all of the information content of the signal is captured in $N =P[(2L-1)(L-1)+1] \sim 2 P L^2$ samples on the ball. The three-dimensional sampling consists of spherical shells, discretised according to the MW sampling theorem, located at the nodes of the radial sampling. The radial sampling may furthermore be rescaled to any spherical region of interest $[0, R] \times \sphere$ using the parameter $\tau$ to dilate or contract the radial quadrature rule.

We have developed the public {\tt FLAG}\footnote{\url{http://www.flaglets.org/}} code \cite{leistedt:flaglets} to compute the Fourier-Laguerre transform. The {\tt FLAG} code computes exact forward and inverse Fourier-Laguerre transforms at machine precision and is stable to extremely large band-limits, relying on the public {\tt SSHT}\footnote{\url{http://www.spinsht.org/}} code \cite{mcewen:fssht} developed by one of the authors for the angular part, which in turn relies on {\tt FFTW}\footnote{\url{http://www.fftw.org/}}.  {\tt FLAG} supports both the C and Matlab programming languages.

%%%%%%%%%%%%%%%%%%%%%%%%%%%%%%%%%%%%%%%%%%%%%%%%%%%%%%%%%%%%%
\subsection{Wavelets on the ball}\label{sec:waveletsball}

To construct wavelets on the ball, we make use of a convolution operator on the ball, which, by separability of the Fourier-Laguerre transform, is derived from convolution operators on the sphere and radial half-line\cite{leistedt:flaglets}. On the sphere, we adopt the usual convolution of $f \in L^2(\sphere)$, detailed in \eg\  Wiaux et al.\cite{wiaux2008dirwavelets}, with an axisymmetric kernel $h \in L^2(\sphere)$ given by the inner product 
\begin{equation}
	(f \star h)(\sas) \equiv \langle f | \mathcal{R}_{(\sas)} h \rangle_{\sphere} = \int_{\sphere} {\rm d}\Omega(\sasprime) f(\sasprime) \left( \mathcal{R}_{(\sas)} h \right)^*(\sasprime).
\end{equation}
where ${\rm d}\Omega(\sas) = \sin\theta {\rm d}\theta {\rm d}\phi$ is the usual rotation invariant measure on the sphere. The translation operator on the sphere is given by the standard three-dimensional rotation: $(\mathcal{R}_{(\alpha,\beta, \gamma)} h)(\sas) = h(\mathcal{R}_{(\alpha,\beta, \gamma)}^{-1}(\sas))$, with $(\alpha,\beta, \gamma) \in {\rm SO(3)}$, where $\alpha\in[0,2\pi)$, $\beta \in[0,\pi]$ and $\gamma \in[0,2\pi)$.  We make the association $\theta=\beta$ and $\phi=\alpha$, \ie\ $\mathcal{R}_{(\sas)} \equiv \mathcal{R}_{(\alpha,\beta,0)}$, and restrict our attention to convolution with axisymmetric functions that are invariant under azimuthal rotation, \ie\ $\mathcal{R}_{(0,0,\gamma)} h = h$, so that we recover a convolved function $f \star h$ defined on the sphere. In harmonic space, axisymmetric convolution may be written 
\begin{equation}
	{(f \star h)}_{\ell m} = \langle f \star h | Y_{\ell m} \rangle_{\sphere} =  \sqrt{ \frac{4\pi}{2\ell+1}} {f}_{\ell m} {h}^*_{\ell 0},
\end{equation}
with $f_{\ell m} = \langle f|Y_{\ell m} \rangle_{\sphere}$ and $h_{\ell 0}\delta_{m 0} = \langle h|Y_{\ell m} \rangle_{\sphere}$. 

On the radial half-line, we adopt a convolution similar to that considered by Gorlich\cite{gorlich:1982} and others (see additional references contained in Ref.~\citenum{gorlich:1982}). We define a translation operator $\mathcal{T}$ by its action on the spherical Laguerre basis functions:
\begin{equation}
	(\mathcal{T}_s K_p )(r) \equiv K_p(s) K_p(r), \label{laguprodconv}
\end{equation}
where $s \in \mathbb{R}^{+}$ (since $K_p$ is real we drop the
complex conjugation).  This leads to a natural harmonic expression for the translation of a radial function $f\in L^2(\mathbb{R}^+)$:
\begin{equation}
  \label{eqn:translation_radial_harmonic_expansion} (\mathcal{T}_{s}f)(r) = \sum_{p = 0}^{\infty} f_{p}  K_p(s) K_{p} (r) ,
\end{equation}
implying $ {(\mathcal{T}_{s}f)}_p = K_p(s) f_{p} $,  where  ${f}_p = \langle f| K_p \rangle_{\mathbb{R}^+}$.  This approach is analogous with the case for the infinite line, for which the standard orthogonal basis is given by the complex exponentials $\phi_\omega(x) = \exp{{\rm i} \omega x}$, with $x,\omega \in \mathbb{R}$.  Translation of the basis functions on the infinite line is simply defined by the shift of coordinates:
$ (\mathcal{T}_{u}^\mathbb{R} \phi_\omega)(x)  \equiv \phi_\omega(x-u) = \phi_\omega^*(u) \phi_\omega(x),$  
with $u\in\mathbb{R}$ and where the final equality follows by the standard rules for exponents. 

It was shown in Ref.~\citenum{mcewen:flagletssampta} that the translation of a function in real space using the radial translation operator is equivalent to its convolution with the shifted Dirac delta function:
\begin{equation}
	(\mathcal{T}_{s}f)(r) = (f \star \delta_s)(r).
\end{equation}
The Dirac delta on the radial half-line at position $s$ is defined as $\delta_s \equiv r^{-2}\delta^{\mathbb{R}}$, where $\delta^{\mathbb{R}}$ is the usual Dirac delta on the infinite line $\mathbb{R}$, such that its harmonic expansion reads $\delta_s(r) = \sum_{p=0}^\infty K_p(s) K_p(r)$. It satisfies the following normalisation and sifting properties, respectively,
\begin{equation}
	\int_{\mathbb{R}^+} {\rm d}r r^2 \delta_s(r) = 1 \quad {\rm and} \quad \int_{\mathbb{R}^+} {\rm d}r r^2 f(r) \delta_s(r) = f(s).
\end{equation}

\begin{figure}\centering
\setlength{\unitlength}{.5in}
\begin{minipage}{8.4cm}
\subfigure[tiling of the Fourier-Laguerre harmonic space]{\begin{picture}(9,7)(0,0)
\put(0.9,0.){\includegraphics[trim = 10.9cm 0cm 1cm 0cm, width=7.5cm]{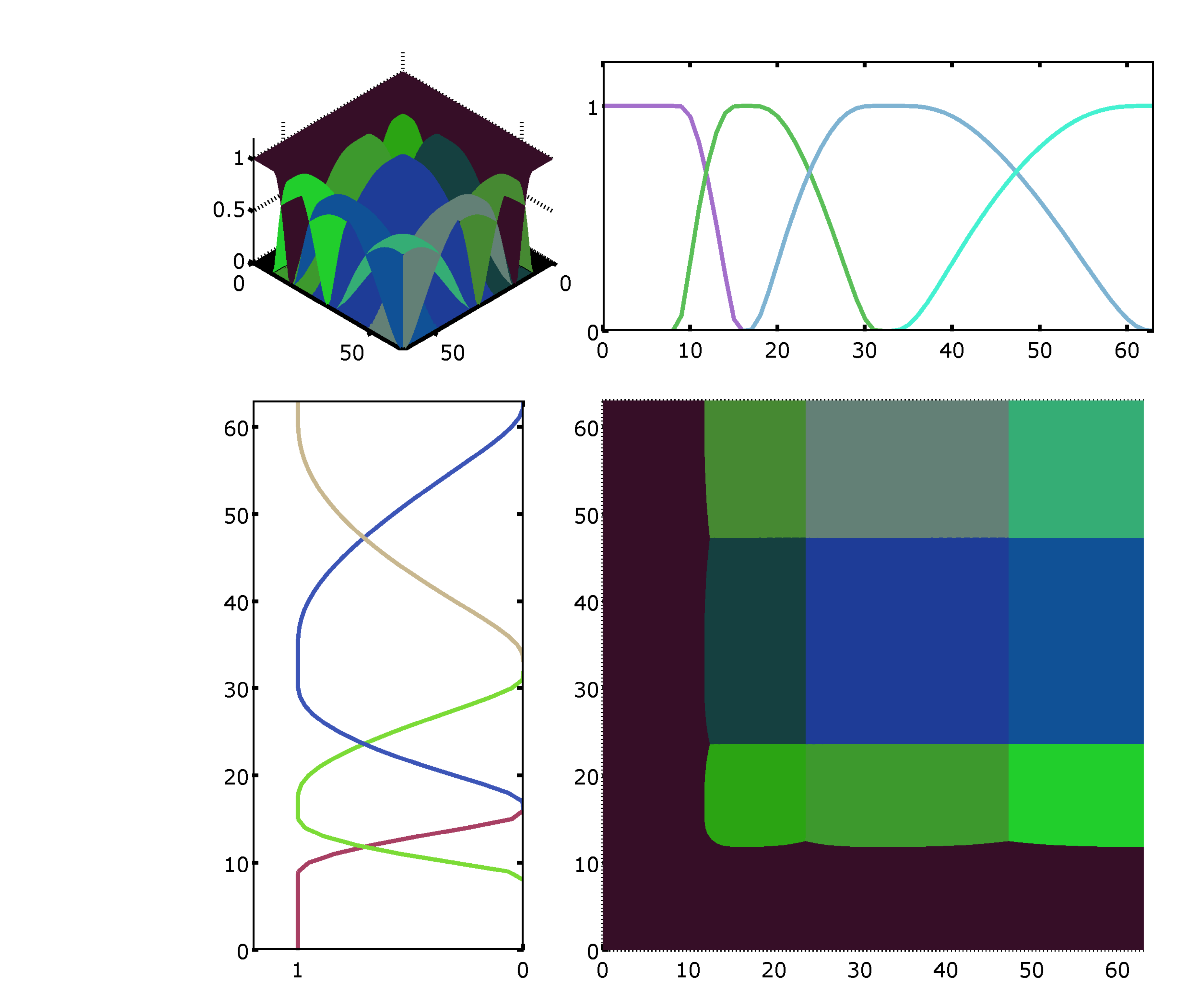}}
\put(0.7,4.7){\scriptsize$\ell$}
\put(2.3,4.7){\scriptsize$p$}
\put(0.13,2.2){\scriptsize$\ell$}
\put(2.5,2.2){\scriptsize$\ell$}
\put(4.87,0){\scriptsize$p$}
\put(4.87,4.32){\scriptsize$p$}
\end{picture}}
\end{minipage} 
\begin{minipage}{8.4cm}
\subfigure[Flaglet translated by $r=0.2$]{\includegraphics[trim= 1cm 24.4cm 0.5cm 1cm, clip, width=4cm]{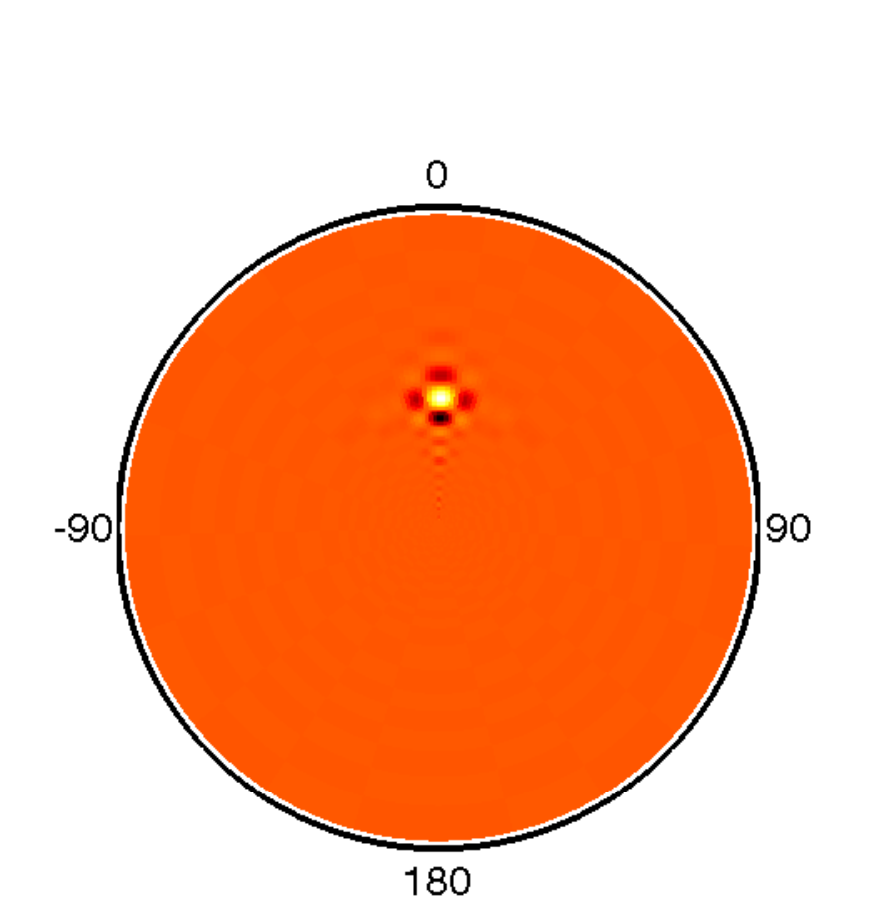}\includegraphics[trim= 6cm 23.4cm 1cm 1cm, clip, width=4cm]{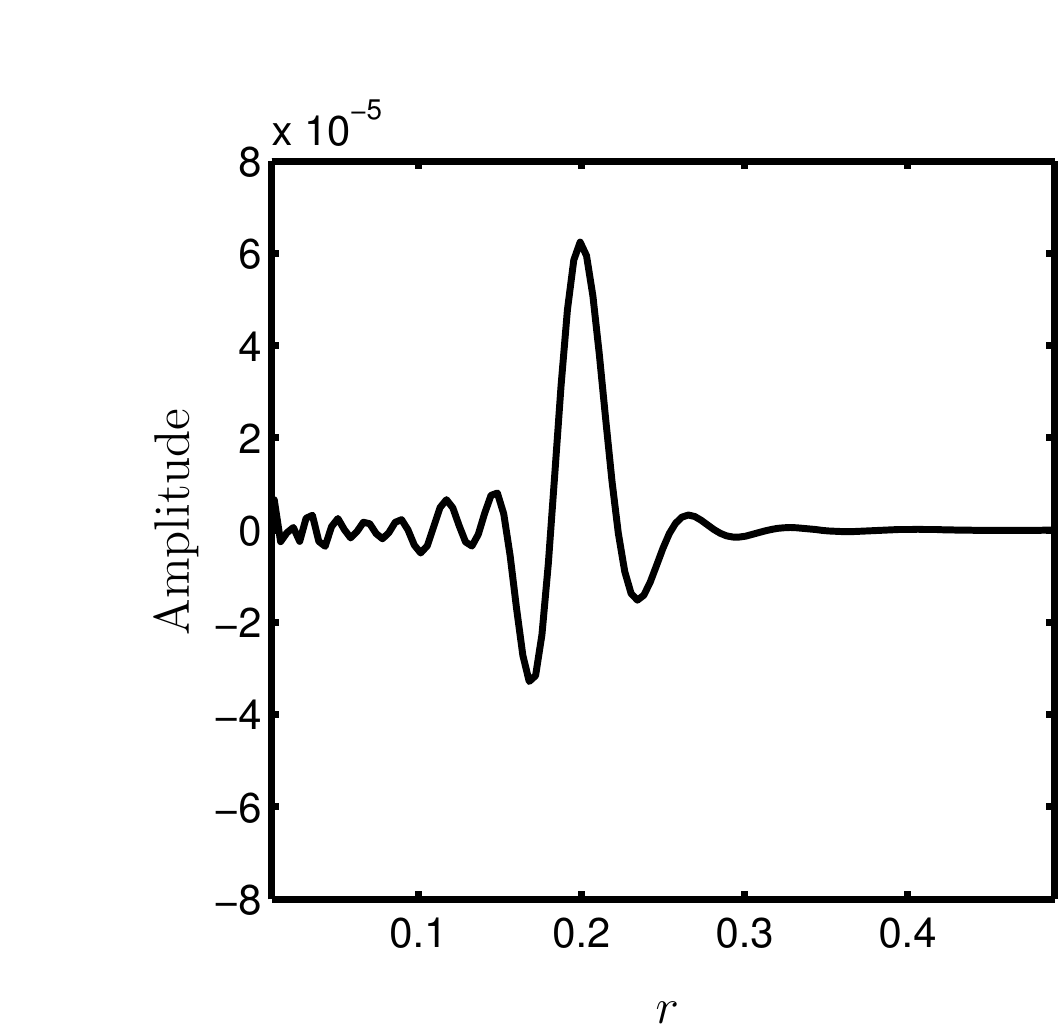}} \\\subfigure[Flaglet translated by $r=0.3$]{\includegraphics[trim= 1cm 14cm 0.5cm 12.5cm, clip, width=4cm]{pics/translationoperator_bis.pdf}\includegraphics[trim= 6cm 12.9cm 1cm 12.4cm, clip, width=4cm]{pics/translationoperator.pdf}}
\end{minipage}
\caption{(a) Tiling of Fourier-Laguerre space at resolution $L=N=64$ for flaglet parameters $\lambda=\nu=2$, giving $J=J^\prime=7$. Flaglets divide {Fourier-Laguerre} space into regions corresponding to specific scales in angular and radial space. The scaling part, here chosen as $J_0=J^\prime_0=4$, is introduced to cover the low frequency region and insures that large scales are also represented by the transform. \quad (b) Slices of an axisymmetric flaglet wavelet ($j=j^\prime=5$) resulting from this tiling, constructed on the ball of radius $R=1$ and translated along the radial half-line. For clarity we zoomed on the range $r\in[0,0.5]$ (the slice hence relates to a ball of radius $r=0.5$). The three-dimensional wavelet can be visualised by rotating this slice around the vertical axis passing through the origin. The translation on the radial half-line not only translates the main feature (the wavelet peak) but also accounts for a damping factor. } 
\label{fig:tiling}
\end{figure}

We define the convolution of two functions on the ball $f, h \in L^2(\ball)$, where $h$ is again assumed to be axisymmetric in the angular direction, by combining the convolution operators defined on the sphere and radial half-line, yielding
\begin{eqnarray}
	(f \star h)(\vect{r}) &\equiv& \langle f | \mathcal{T}_r  \mathcal{R}_{(\sas)} h \rangle_{\ball}  \ = \ \int_{\ball} {\rm d}^3\vect{r}^\prime f(\vect{r}^\prime) \left( \mathcal{T}_r \mathcal{R}_{(\sas)} h \right)^*(\vect{r}^\prime),
\end{eqnarray}
given in harmonic space by the product
\begin{equation}
	{(f \star h)}_{\ell m p}=  \langle f \star h | Z_{\ell m p} \rangle_{\ball} =  \sqrt{ \frac{4\pi}{2\ell+1}} {f}_{\ell mp} {h}^*_{\ell 0 p} ,
\end{equation}
with $f_{\ell m p} = \langle f|Z_{\ell m p} \rangle_{\ball}$ and $h_{\ell 0 p}\delta_{m 0} = \langle h|Z_{\ell mp} \rangle_{\ball}$. The action of the translation operator is illustrated in \fig{\ref{fig:tiling}} through its action on the wavelet functions defined below.

With an exact harmonic transform and a convolution operator in hand, we are now in a position to construct the exact flaglet transform on the ball. For a function of interest $ f\in L^2(\ball)$, we define its $jj^\prime$-th wavelet coefficient $W^{\Psi^{jj^\prime}}\in L^2(\ball)$ as the convolution of $f$ with the wavelet $\Psi^{jj^\prime}\in L^2(\ball)$:
\begin{equation}
 	W^{\Psi^{jj^\prime}}(\vect{r})  \equiv (f \star \Psi^{jj^\prime})(\vect{r}) = \langle f | \mathcal{T}_r \mathcal{R}_{(\sas)} \Psi^{jj^\prime} \rangle_{\ball}. \label{wav1}
\end{equation} 
The scales $j$ and $j^\prime$ respectively relate to angular and radial spaces. Since we restrict ourselves to axisymmetric wavelets, the wavelet coefficients are given in {Fourier-Laguerre} space by the product
\begin{equation}
	{W}^{\Psi^{jj^\prime}}_{\ell m p} =  \sqrt{ \frac{4\pi}{2\ell+1}} {f}_{\ell m p} {\Psi}^{jj^\prime*}_{\ell 0 p}, \label{wav2}
\end{equation}
where ${W}^{\Psi^{jj^\prime}}_{\ell m p} = \langle W^{\Psi^{jj^\prime}} |Z_{\ell m p} \rangle_{\ball}$, $f_{\ell m p} = \langle f|Z_{\ell m p} \rangle_{\ball}$ and ${\Psi}^{jj^\prime}_{\ell 0 p}\delta_{m 0} = \langle \Psi^{jj^\prime}|Z_{\ell m p} \rangle_{\ball}$.  The wavelet coefficients contain the detail information of the signal only; a scaling function and corresponding scaling coefficients must be introduced to represent the low-frequency, approximate information of the signal.  The scaling coefficients $W^\Phi \in L^2(\ball)$ are defined by the convolution of $f$ with the scaling function $\Phi\in L^2(\ball)$:
\begin{equation}
W^\Phi(\vect{r})  \equiv (f \star \Phi)(\vect{r}) = \langle f | \mathcal{T}_r \mathcal{R}_{(\sas)} \Phi \rangle, \label{wav3}
\end{equation}
or in {Fourier-Laguerre} space,
\begin{equation}
	{W}^\Phi_{\ell m p} =  \sqrt{ \frac{4\pi}{2\ell+1}} {f}_{\ell m p} {\Phi}^{*}_{\ell 0 p}, \label{wav4}
\end{equation}
where ${W}^{\Phi}_{\ell m p} = \langle W^{\Phi} |Z_{\ell m p} \rangle_{\ball}$ and ${\Phi}_{\ell 0 p}\delta_{m 0} = \langle \Phi|Z_{\ell m p} \rangle_{\ball}$. 

Provided the wavelets and scaling function satisfy an admissibility property, a function $f$ may be reconstructed exactly from its wavelet and scaling coefficients by
\begin{eqnarray}
	\quad f(\vect{r}) \ \ = \ \ \int_{\ball} {\rm d}^3\vect{r}^\prime W^{\Phi}(\vect{r}^\prime)(\mathcal{T}_r \mathcal{R}_{(\sas)} \Phi)(\vect{r}^\prime) \ + \ \sum_{j=J_0}^{J} \sum_{j^\prime=J^\prime_0}^{J^\prime}  \int_{\ball} {\rm d}^3\vect{r}^\prime W^{\Psi^{jj^\prime}}\hspace{-2mm}(\vect{r}^\prime)(\mathcal{T}_r \mathcal{R}_{(\sas)} \Psi^{jj^\prime})(\vect{r}^\prime), 
\end{eqnarray}
or equivalently in harmonic space by
\begin{eqnarray}
	 {f}_{\ell m p} &=& \sqrt{ \frac{4\pi}{2\ell+1}} {W}^{\Phi}_{\ell m p} {\Phi}_{\ell 0 p}   
	 \ + \ \sqrt{ \frac{4\pi}{2\ell+1}} \sum_{j=J_0}^{J} \sum_{j^\prime=J^\prime_0}^{J^\prime}   {W}^{\Psi^{jj^\prime}}_{\ell m p} {\Psi}^{jj^\prime}_{\ell 0 p}. \label{waveletsynthesis}
\end{eqnarray}
The parameters $J_0$, $J^\prime_0$, $J$ and $J^\prime$ defining the minimum and maximum scales must be defined consistently to extract and reconstruct all the information contained in $f$. They depend on the construction of the wavelets and scaling function and are defined explicitly in the next paragraph. 

Finally, the admissibility condition under which a band-limited function $f$ can be decomposed and reconstructed exactly is given by the following resolution of the identity:
\begin{equation}
	\frac{4\pi}{2\ell+1} \left( |{\Phi}_{\ell 0 p}|^2 + \sum_{j=J_0}^{J} \sum_{j^\prime=J^\prime_0}^{J^\prime}  |{\Psi}^{jj^\prime}_{\ell 0 p}|^2 \right) \ = \ 1, \quad \forall \ell, p . \label{identity}
\end{equation}

We may now construct wavelets and scaling functions that satisfy this admissibility property and thus lead to an exact wavelet transform on the ball.  We extend the notion of harmonic tiling \cite{Marinucci2008needlets, wiaux2008dirwavelets, 2010needatool, leistedt:s2let_axisym} to the Fourier-Laguerre space and construct axisymmetric wavelets (flaglets) {well localised in both real and {Fourier-Laguerre} spaces}.  We first define the flaglet and scaling function generating functions, before defining the flaglets and scaling function themselves.

We start by considering the $C^{\infty}$ Schwartz function with real parameter $\lambda\in\mathbb{R}^+_*$ and compact support $[\frac{1}{\lambda}, 1]$
\begin{equation}
	s_\lambda(t) \equiv s\left( \frac{2\lambda}{\lambda-1} (t-1/\lambda)-1\right) \textrm{\quad with \quad } s(t) \equiv \left\{ \begin{array}{ll} \ e^{-\frac{1}{1-t^2}}, & t\in[-1,1] \\ \  0, & t \notin [-1,1]\end{array} \right. .
\end{equation}
We then define the smoothly decreasing function $k_\lambda$ by
\begin{equation}
	 k_\lambda(t) \equiv \frac{\int_{t}^1\frac{{\rm d}t^\prime}{t^\prime}s_\lambda^2(t^\prime)}{\int_{1/\lambda}^1\frac{{\rm d}t^\prime}{t^\prime}s_\lambda^2(t^\prime)}, \label{smoothscaling}
\end{equation}
which is unity for $t<1/\lambda$, zero for $t>1$, and is smoothly decreasing from unity to zero for $t \in [1/\lambda,1]$. 
Axisymmetric flaglets are constructed in a two-dimensional space corresponding to the harmonic indices $\ell$ and $p$. We associate $\lambda$ with $\ell$-space and we introduce a second parameter $\nu$ associated with $p$-space, with the corresponding functions $k_\lambda$ and $k_\nu$. We define the flaglet generating function by
\begin{equation}
	 \kappa_\lambda(t) \equiv \sqrt{ k_\lambda(t/\lambda) - k_\lambda(t) }
\end{equation}
and the scaling function generating function by
\begin{equation}
	 \eta_{\lambda}(t) \equiv \sqrt{ k_\lambda(t)} ,
\end{equation}
with similar expressions for {$\kappa_\nu$ and $\eta_\nu$}, complemented with a hybrid scaling function generating function
\begin{eqnarray}
	\eta_{\lambda \nu}(t, t^\prime) \equiv  \sqrt{ \ k_\lambda(t/\lambda)k_\nu(t^\prime)   +  k_\lambda(t)k_\nu(t^\prime/\nu)   -   k_\lambda(t)k_\nu(t^\prime) \ }.
\end{eqnarray}

The flaglets and scaling function are constructed from their generating functions to satisfy the admissibility condition given by \eqn{\ref{identity}}. A natural approach is to define ${\Psi}^{jj^\prime}_{\ell m p}$ from the generating functions $\kappa_\lambda$ and $\kappa_\nu$ to have support on $[\lambda^{j-1},\lambda^{j+1}]\times[\nu^{j^\prime-1},\nu^{j^\prime+1}]$, yielding
\begin{equation}
	{\Psi}^{jj^\prime}_{\ell m p} \equiv \sqrt{ \frac{2 \ell+1}{4\pi}}  \ \kappa_\lambda\left(\frac{\ell}{\lambda^j}\right) \kappa_\nu\left(\frac{\phantom{\ell}\hspace*{-2mm}p}{\nu^{j^\prime}}\right) \delta_{m0}.
\end{equation}
With these waveletss, \eqn{\ref{identity}} is satisfied for $\ell > \lambda^{J_0}$ and \mbox{$p > \nu^{J^\prime_0}$}, where $J_0$ and $J^\prime_0$ are the lowest wavelet scales used in the decomposition. The scaling function $\Phi$ is constructed to extract the modes that cannot be probed by the flaglets:\footnote{Note that despite its piecewise definition ${\Phi}_{\ell m p}$ is continuous along and across the boundaries $p=\nu^{J_0^\prime}$ and $\ell = \lambda^{J_0}$. }
\begin{equation}
	{\Phi}_{\ell m p}  \equiv   \left\{ \begin{array}{ll} 
	 \hspace{-1mm} \sqrt{ \frac{2 \ell+1}{4\pi}}  \ \eta_{\nu}\left(\frac{\phantom{\ell}\hspace*{-2mm}p}{\nu^{J^\prime_0}}\right)\delta_{m0} , & \textrm{if } \ell  > \lambda^{J_0}, \ p \leq \nu^{J^\prime_0}	 \\
  \hspace{-1mm} \sqrt{ \frac{2 \ell+1}{4\pi}}  \ \eta_{\lambda}\left(\frac{\ell}{\lambda^{J_0}}\right)\delta_{m0}  , & \textrm{if } \ell \leq \lambda^{J_0} , \ p  > \nu^{J_0^\prime}  \\
\hspace{-1mm}  \sqrt{ \frac{2 \ell+1}{4\pi}}  \ \eta_{\lambda \nu}\left(\frac{\ell}{\lambda^{J_0}},\frac{p}{\nu^{J^\prime_0}}\right)\delta_{m0} ,  \hspace{-2mm} & \textrm{if }  \ell < \lambda^{J_0}, \ p  < \nu^{J_0^\prime} \\
  \quad 0 , & \textrm{elsewhere.} \end{array} \right. 
\end{equation}

\begin{figure}
\centering
\includegraphics[trim=6cm 2cm 4cm 1cm, clip, width=14cm]{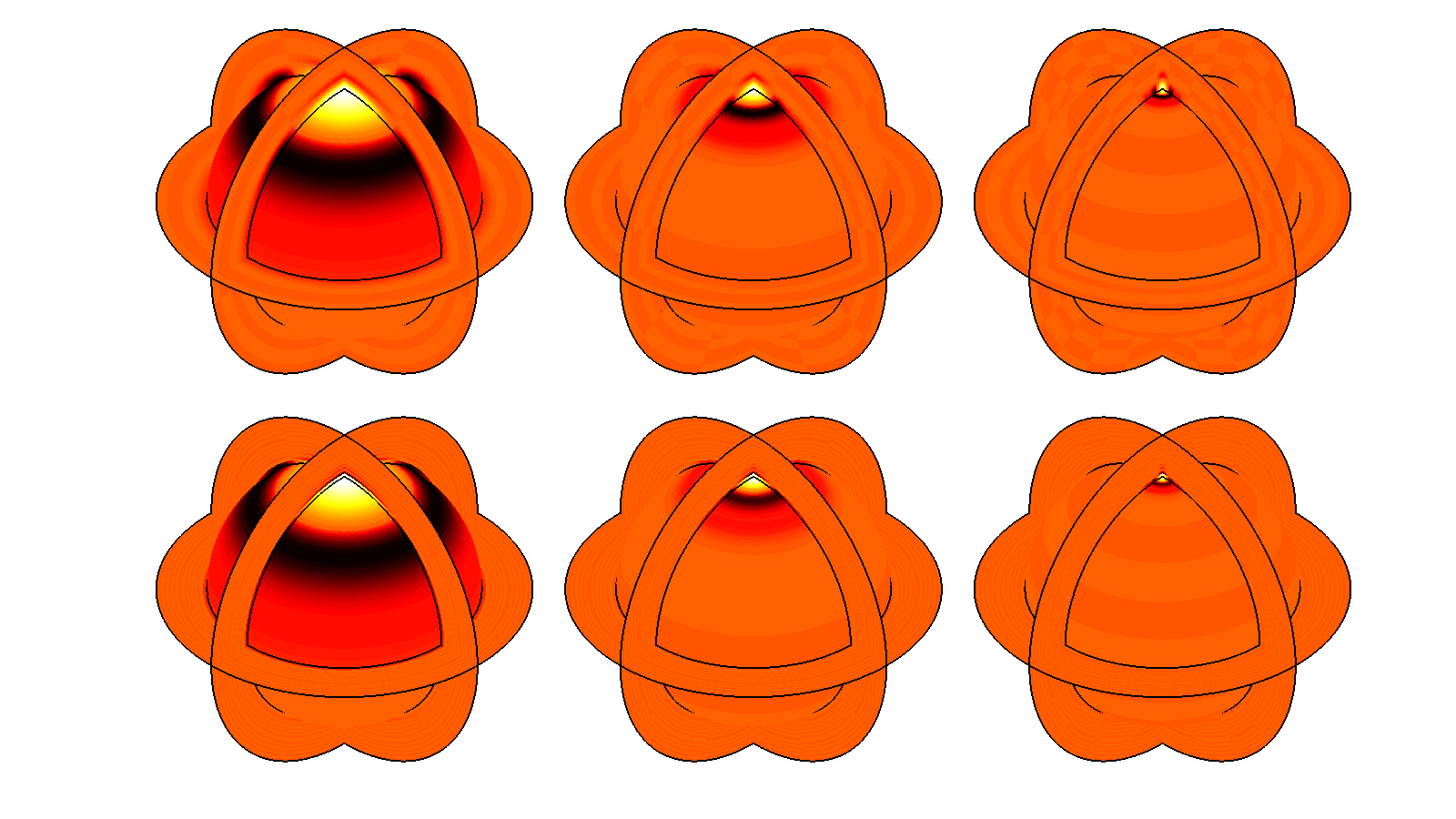}
\caption{Successive axisymmetric flaglet wavelet arising from a tiling of the Fourier-Laguerre harmonic space with $\lambda=\nu=3$. For visualisation purposes we show the flaglets corresponding to $j\in\{2,3,4\}$ and $j^\prime\in\{4,5\}$, translated to $r=0.25$, reconstructed at resolution $L=N=256$ on a ball of radius $R=0.4$ (with a spherical slice at $r=0.25$). Flaglets of angular order $j=2$ (first column) probe large angular scales compared to those of order $j=3,4$ (second and third columns). Similarly, flaglets of radial order $j^\prime=4$ (first row) probe large radial scales compared to those of order $j^\prime=5$ (second row).
Flaglets are well localised in both real and {Fourier-Laguerre} spaces and their angular (radial) aperture is invariant under radial (angular) translation.}
\label{fig:flaglets}
\end{figure}

To satisfy exact reconstruction, $J$ and $J'$ are defined from the band-limits by $J = \lceil \log_\lambda(L-1) \rceil$ and \mbox{$J^\prime = \lceil \log_\nu(P-1) \rceil$}. The choice of $J_0$ and $J^\prime_0$ is arbitrary, provided that $0 \leq J_0 < J$ and $0 \leq J^\prime_0 < J^\prime$.  This framework generalises the notion of the harmonic tiling used to construct exact wavelets on the sphere \cite{Marinucci2008needlets, wiaux2008dirwavelets}; in fact, the flaglets defined here reduce in angular part to the wavelets defined in Ref.~\citenum{wiaux2008dirwavelets} for the axisymmetric case.  The flaglets and scaling function tiling of the Fourier-Laguerre space of the ball is shown in \fig{\ref{fig:tiling}}. The flaglets resulting from this tiling are presented in \fig{\ref{fig:flaglets}}. Flaglets and the scaling function may be reconstructed in the spatial domain from their harmonic coefficients. The flaglets are well localised in both real and Fourier-Laguerre spaces and their angular (radial) aperture is invariant under radial (angular) translation. They also form a tight frame, as detailed in Ref.~\citenum{mcewen:flagletssampta}.

We have developed the public code {\tt FLAGLET}\footnote{\url{http://www.flaglets.org/}} \cite{leistedt:flaglets} to compute the flaglet transform, relying on the public codes {\tt FLAG}, {\tt S2LET}\footnote{\url{http://www.s2let.org/}}\cite{leistedt:s2let_axisym}, {\tt SSHT}\footnote{\url{http://www.spinsht.org/}} and {\tt FFTW}\footnote{\url{http://www.fftw.org/}}.  {\tt FLAGLET} supports both the C and Matlab programming languages.

%%%%%%%%%%%%%%%%%%%%%%%%%%%%%%%%%%%%%%%%%%%%%%%%%%%%%%%%%%%%%
\subsection{Connection to Fourier-Bessel analysis}\label{sec:fourierbessel}

As an alternative to the spherical Laguerre transform, the spherical Bessel transform is a radial transform based on the eigenfunctions of the Laplacian operator in spherical coordinates. This basis is central to the Fourier-Bessel transform, currently used in cosmology \cite{leistedt20113dex, rassat2011baos, abramo2010cmbbox} to analyse the spectral properties of galaxy surveys in three dimensions. More precisely, the Fourier-Bessel transform of $f \in L^2(\ball)$ reads
\begin{equation}
	f(\vect{r}) = \sum_{\ell = 0}^{\infty}\sum_{m = -\ell}^{\ell} \sqrt{\frac{2}{\pi}} \int_{\mathbb{R}^+} {\rm d}k k^2 \tilde{f}_{\ell m}(k) Y_{\ell m} (\sas) j_\ell(kr), \label{fbinverse}
\end{equation}
where $j_\ell$ is the spherical Bessel function of order $\ell$ the harmonic coefficients are given by the projection
\begin{equation}
	\tilde{f}_{\ell m}(k) = \langle f | Y_{\ell m} j_\ell \rangle_{\ball} =  \sqrt{\frac{2}{\pi}} \int_{\sphere}  {\rm d}\Omega(\sas)  \int_{\mathbb{R}^+} {\rm d}r r^2  f(r, \sas) Y^*_{\ell m}(\sas) j_\ell(kr).  \label{fbinverse}
\end{equation}
 
Compared to the Fourier-Laguerre transform, the radial and angular components of the Fourier-Bessel basis are not fully separable since they share the $\ell$ mode index. Moreover, to our knowledge, there exists no method to compute the integral of \eqn{\ref{fbinverse}} exactly for a useful class of functions. The origin of this issue is the difficulty of finding a quadrature formula for the spherical Bessel functions on $\mathbb{R}^+$. Finally, the use of numerical integration methods does not always guarantee good accuracy because of the oscillatory nature of the spherical Bessel functions. 

In Ref.~\citenum{leistedt:flaglets} we derived an analytical formula to compute the Fourier-Bessel transform of a signal from its Fourier-Laguerre transform:
\begin{equation} 
	\tilde{f}_{\ell m}(k) = \sqrt{\frac{2}{\pi}} \sum_p {f}_{\ell m p}  {j}_{\ell p}(k). \label{exactbessel}
\end{equation}
This summation is finite if the signal is band-limited in Fourier-Laguerre space, in which case this approach is exact, \ie\ gives access to the Fourier-Bessel coefficients of $f$ without the need of a sampling theorem for the Fourier-Bessel transform. In this formula, the functions ${j}_{\ell p} (k)$ are the projections of the spherical Laguerre basis functions on the spherical Bessel functions. Using the properties of the bases, the projections reduces to
 \begin{equation}
 	{j}_{\ell p} (k) \equiv \langle K_p | j_\ell \rangle =   \int_{\mathbb{R}^+} {\rm d} r r^2 K_p(r)  j^*_\ell(k r)  =  \sqrt{ \frac{p!}{(p+2)!} }  \sum_{j=0}^{p}   c^{p}_{j} \mu^\ell_{j+2} (k), \label{jlpk} 
\end{equation}
where the $c^p_{j}$ satisfy the following recurrence:
\begin{equation}
	c^{p}_{j} \equiv \frac{(-1)^j}{j!} {p+2 \choose p - j} =  -\frac{p-j+1}{j(j+2)} c^{p}_{j-1}.
\end{equation}%
The functions $\mu^\ell_{j} (k)$ are the moments of $j_\ell(kr)e^{-\frac{r}{2\tau}}$, \ie 
\begin{equation}
	\mu^\ell_{j} (k) \equiv \frac{1}{\tau^{j-\frac{1}{2}}} \int_{\mathbb{R}^+} {\rm d} r r^j j_\ell(kr) e^{-\frac{r}{2\tau}}.
\end{equation}
Watson \cite{watson1995treatise} derived an analytical solution for the latter integral:
\begin{eqnarray}\small
	\quad \mu^\ell_j(k)   &=& \ \sqrt{\pi} \ 2^{j}\ \tilde{k}^{\ell} \ \tau^\frac{3}{2}  \ \frac{  \Gamma(j + \ell + 1) }{ \Gamma(\ell+\frac{3}{2}) }  \label{muljk}  \ \hspace{-2mm} \phantom{s}_2F_1 \left( \frac{j + \ell + 1}{2}  ;  \frac{ j + \ell}{2} + 1 ; \ell+\frac{3}{2} ; -4\tilde{k}^2 \right)
\end{eqnarray}
where $\tilde{k} = \tau k$ is the rescaled $k$ scale and$\phantom{s}_2F_1$ is the Gaussian hypergeometric function. Since either $(j+\ell+1)/2$ or $(j+\ell)/2$ is a positive integer, the hypergeometric function reduces to a polynomial of $\tilde{k}^2$ and it is possible to compute the quantity ${j}_{\ell p} (k)$ exactly using \eqn{\ref{jlpk}} to \eqn{\ref{muljk}}. Consequently, the inverse Fourier-Bessel transform $\tilde{f}_\ell(k)$ may be calculated analytically through {\ref{exactbessel}}, which is exact if $f$ is band-limited in the Fourier-Laguerre basis.

The Fourier-Bessel transform is also suitable for a wavelet construction. Lanusse et al.\cite{mrs3d} defined an isotropic wavelet 3D transform on the Fourier-Bessel basis by extending the framework of isotropic spherical wavelets introduced by Stark et al.\cite{starck2006ridge}. Since these wavelets are isotropic in three dimensions, their angular aperture depends on the distance to the origin, which may not be ideal to treat the angular and radial selection effects and uncertainties in galaxy survey data. In addition, the definitions of the isotropic wavelet transform in the continuous and discrete settings are not equivalent (\ie\ the transform is not theoretically exact) due to the absence of an exact quadrature formula for the spherical Bessel transform (the radial part of the Fourier-Bessel transform) \cite{lemoine1994sbt}. Nevertheless, this transform achieves good numerical accuracy and is useful to probe spatially localised isotropic features in signals on the ball.

%%%%%%%%%%%%%%%%%%%%%%%%%%%%%%%%%%%%%%%%%%%%%%%%%%%%%%%%%%%%%
\section{Cosmic voids: a potential application}\label{sec:voids}
 
Galaxy surveys collect the positions and properties of galaxies on the sky in order to map the three-dimensional large-scale structure of the Universe. Interestingly, these galaxies are not observed in three dimensions directly: their angular positions are enhanced with redshift information estimated from the spectra of their light. This setting yields data with a spherical geometry, distorted by complex selection effects and systematic uncertainties which are naturally defined on the sky and along the line of sight.
These contaminants for example arise from the uncertainties in the redshift estimates and the variations in the magnitude limits, observing conditions and calibration of the instruments over time. They severely complicate the three-dimensional analysis of galaxy survey data. To overcome this issue, one can opt for a two-dimensional spherical analysis, and group galaxies in large redshift bins to study certain properties of interest (\eg\  their clustering) projected on the sky. In this case, standard techniques on the sphere can be used, such as the spherical harmonic transform, to confront the data with theoretical predictions from cosmological models. However, the 3D distribution of galaxies contains a tangible amount of information which is underexploited by these projected angular analyses. Full three-dimensional approaches have been developed previously and used to exploit galaxy surveys. For instance, the Fourier-Bessel transform, the basis functions of which are eigenfunctions of Laplacian operator in 3D spherical coordinates, was previously used to reconstruct the density and velocity fields of the local Universe using galaxy survey data\cite{erdogdu20046df, erdogdu20062mass}. However, all current 3D approaches have non-separable angular and radial components, which complicates the treatment of the survey geometry, uncertainties and selection effects, naturally given on the sky and along the line of sight separately. The Fourier-Laguerre basis can address this problem since it is appropriate for a separable harmonic analysis in 3D spherical coordinates. Furthermore, clusters of galaxies, filaments and voids in the large-scale structure are spatially localised and therefore can potentially be extracted using flaglets. In this paper, we investigate the use of flaglets for detecting cosmic voids in galaxy surveys. This application will be fully developed in future work, and is presented as a proof of concept here. 

Cosmic voids are the large, underdense regions that occupy a significant fraction of the volume of the Universe and are a natural consequence of the hierarchical growth of structure\cite{Weygaert2011voiddynamics, Aragon2013hierach,Sheth2004hierach,Colberg2005lcdmvoids}. They trace the large-scale structure just like galaxies do, and have a high potential for learning about the properties of dark matter and distinguishing between models of dark energy\cite{BLi2013modgravityvoids,Sutter2012alcockvoids,Biswas2010voidsdarkenergy}. In particular, unlike galaxies which are affected by non-linear gravitational interactions, voids are in the well-understood quasi-linear regime, which simplifies their modelling when confronting data to theoretical predictions. In addition, voids are numerous in the distribution of galaxies, and their statistical power can be exploited to obtain high signal-to-noise ratio in cosmological analyses. However, there is currently no consensus on the best approach to define or find voids in galaxy survey data, due to the presence of complex selection effects and systematic uncertainties. In theory, voids are wells in the gravitational potential, \ie\ underdense regions from which galaxies move away\cite{Lavaux2010voidcosmology, Aragon2013hierach}. This definition is very difficult to apply in practice since it requires reconstructing the gravitational potential, which relies on numerous assumptions about the distribution of dark matter and galaxies, and is also very sensitive to uncertainties in the data. More robust approaches are topological, and define cosmic voids as empty spaces between galaxies. In this context, several techniques exist to identify voids in galaxy surveys, such as watershed-based algorithms\cite{Neyrinck2008zobov, Sutter2012zobovdr7}. However, these topological methods exploit the positions of individual galaxies, even on large scales, and are therefore also sensitive to selection effects and systematic uncertainties.  With an appropriate algorithm, flaglets may overcome these issues since they can probe underdense regions at several scales and be robust to the positions of individual galaxies. In particular, in a 3D map of galaxies counted in voxels (\eg\  on the Fourier-Laguerre sampling), the observational distortions and positions of individual galaxies will only affect the response of the flaglet transform on a limited number of scales and to a smaller extent than pure topological void-finders. 

\begin{figure}
\centering
\subfigure[N-body data]{\includegraphics[trim=6cm 2cm 4cm 1cm, clip, width=8cm]{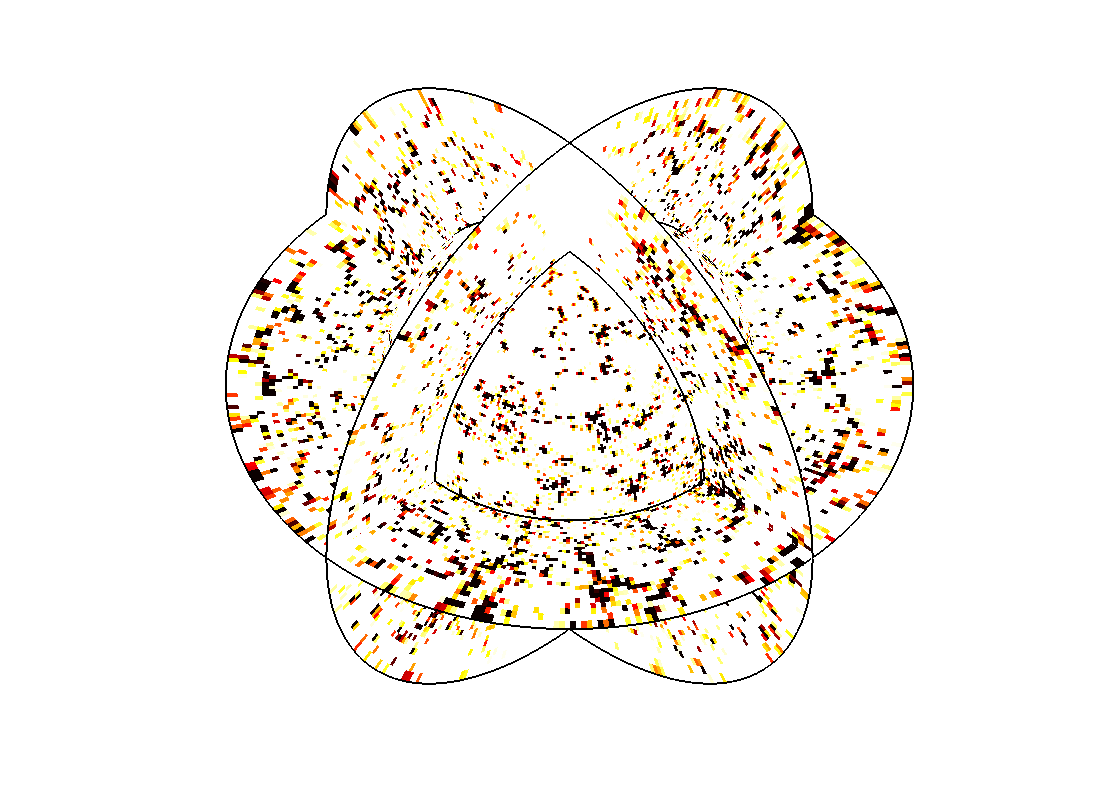}}
\subfigure[Flaglets with negative colour scheme to highlight underdense regions and voids.]{\includegraphics[trim=6cm 2cm 4cm 1cm, clip, width=14cm]{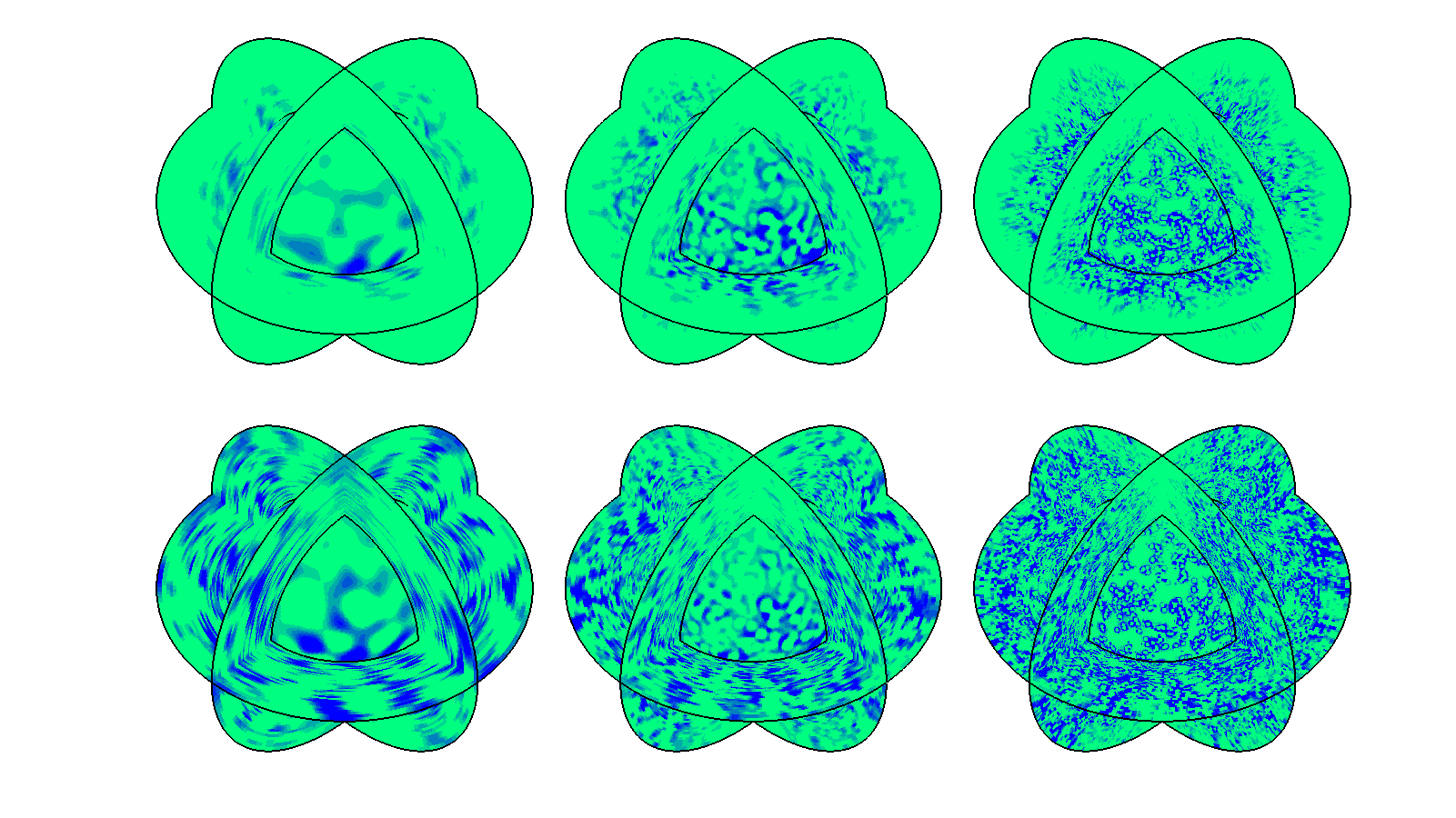}}
\caption{Flaglet decomposition of the N-body simulation dataset, pixelised and band-limited at $L=P=192$. The flaglet parameters are $\lambda=\nu=3$ and the scaling coefficients correspond to $J_0=2, J_0^\prime=4$, since the lower scale indices do not contain a great deal of information. The colour scale was truncated such that blue indicates negative response to the flaglets, \ie\ underdense regions and voids in the galaxy distribution, and green corresponds to zero or positive response, \ie\ filaments and other structures, which are not highlighted here. }
\label{fig:horizonvoids}
\end{figure}

To illustrate this approach, we consider the full-sky Horizon simulation \cite{teyssier2009horizon}: an N-body simulation covering a 1 Gpc periodic box of 70 billion dark matter particles generated from the concordance model cosmology derived from 3-year Wilkinson Microwave Anisotropy Probe (WMAP) observations \cite{wmap3Spergel}. The purpose of such a simulation is to reproduce the action of gravity (and to a minor extent galaxy formation) on a large system of particles, with the initial conditions drawn from a cosmological model of interest. The outcome is commonly used to confront astrophysical models with observations. For simplicity we only consider a ball of 1 MPc radius centered at the origin so that the structures are of reasonable size. \fig{\ref{fig:horizonvoids}} (a) shows the initial data, band-limited at $L=P=192$. It exhibits a filamentary structure, with clusters, sheets and voids at specific physical scales, characteristics of the cosmology under consideration. \fig{\ref{fig:horizonvoids}} (b) shows the flaglet decomposition of these data, with convenient choices for the wavelet parameters and the colour scales to only visualise underdense regions and highlight the ability of flaglets to extract cosmic voids. We used $\lambda = \nu = 3$, and $J_0=2, J_0^\prime=4$ since the lower-scale indices (corresponding to very large scales in the data) do not contain a great deal of information. The colour scale was truncated such that blue indicates negative response to the flaglets, \ie\ voids in the galaxy distribution, and green corresponds to zero or positive response, \ie\ filaments and other structures, which do not interest us here.  We see that the flaglets naturally detect voids at different scales, and that complex selection effects will only change the response of the flaglets over certain scales on the sky and along the line of sight separately. In future work we will design a robust algorithm for locating voids using this approach. In particular, we will exploit flaglets at similar scales to evaluate the significance of  individual voids, and flaglets at very different scales to separate large parent voids from their sub-voids. This approach opens new perspectives for recovering the hierarchical void distribution in the large-scale structure. Naturally, real data will require dealing with selection effects and uncertainties, but these complications are mostly separable on the sky and along the line of sight, and therefore naturally handled in the Fourier-Laguerre basis.

%%%%%%%%%%%%%%%%%%%%%%%%%%%%%%%%%%%%%%%%%%%%%%%%%%%%%%%%%%%%%
\section{Summary}\label{sec:summary}
%%%%%%%%%%%%%%%%%%%%%%%%%%%%%%%%%%%%%%%%%%%%%%%%%%%%%%%%%%%%%

We have reviewed the Fourier-Laguerre transform, a novel 3D spherical transform which combines Laguerre polynomials on the radial half-line with the spherical harmonics on the sphere and benefits from a sampling theorem on the ball. For radial and angular band-limits $P$ and $L$, respectively, the sampling theorem guarantees that all the information of the band-limited signal is captured in a finite set of $N=P[(2L-1)(L-1)+1]$ samples on the ball. Furthermore, the Fourier-Bessel transform of the band-limited signal can be calculated exactly thanks to an analytical connection to the Fourier-Laguerre transform. We have developed an exact wavelet transform on the ball, the flaglet transform, through a tiling of the Fourier-Laguerre space. Flaglets {form a tight frame, are well localised in both real and {Fourier-Laguerre} spaces, and their angular (radial) aperture is invariant under radial (angular) translation.}  Our implementation of all of the transforms detailed in this article is made publicly available. In future work we intend to revoke the axisymmetric constraint by developing directional flaglets.

We have highlighted the efficiency of flaglets for extracting scale-dependent, spatially localised features in data on the ball by considering the problem of finding voids in an N-body simulation. This application will be fully developed in future work, with an emphasis on an accurate treatment of selection effects and uncertainties in galaxy survey data. This approach opens new perspectives for finding and studying voids in galaxy surveys and using their distribution to probe the large-scale structure of the Universe.

%%%%%%%%%%%%%%%%%%%%%%%%%%%%%%%%%%%%%%%%%%%%%%%%%%%%%%%%%%%%%
\acknowledgments     %>>>> equivalent to \section*{ACKNOWLEDGMENTS}       
 
We thank Paul Sutter for useful discussions. BL is supported by the Perren Fund, the IMPACT Fund, and a travel grant by the Royal Astronomical Society to attend SPIE Optics and Photonics 2013.  HVP is supported by STFC, the Leverhulme Trust, and the European Research Council under the European Community's Seventh Framework Programme (FP7/2007- 2013) / ERC grant agreement no 306478-CosmicDawn.  JDM is supported in part by a Newton International Fellowship from the Royal Society and the British Academy.

%%%%%%%%%%%%%%%%%%%%%%%%%%%%%%%%%%%%%%%%%%%%%%%%%%%%%%%%%%%%%
%%%%% References %%%%%

\bibliography{biblio}   %>>>> bibliography data in report.bib
\bibliographystyle{spiebib}   %>>>> makes bibtex use spiebib.bst

\end{document}